\documentclass[10pt,a4paper,twocolumn]{article}
\usepackage{graphicx}

\begin{document}
\title{The Origin of the Planck's Constant}
\author{\small H. Y. Cui\footnote{E-mail: hycui@public.fhnet.cn.net}\\
\small Department of Applied Physics\\
\small Beijing University of Aeronautics and Astronautics\\
\small Beijing, 100083, China}
\date{\small \today}

\maketitle

\begin{abstract}
\small
In this paper, we discuss an equation which does not contain the Planck's
constant, but it will turn out the Planck's constant when we apply the
equation to the problems of particle diffraction.\\

\end{abstract}

\section{Introduction}

In 1900, M. Planck assumed that the energy of a harmornic oscilator can take
on only discrete values which are integral multiples of $h\nu $, where $\nu $
is the vibration frequency and $h$ is a fundamental constant, now either $h$
or $\hbar =h/2\pi $ is called as Planck's constant. The Planck's constant
next made its appearance in 1905, when Einstein used it to explain the
photoelectric effect, he assumed that the energy in an electromagnetic wave
of frequency $\omega $ is in the form of discrete quanta (photons) each of
which has an energy $\hbar \omega $ in accordance with Planck's assumption.
From then, it has been recognized that the Planck's constant plays a key
role in the quantum mechanics.

In this paper, we discuss an equation which does not contain the Planck's
constant, but it will turn out the Planck's constant when we apply the
equation to the problems of particle diffraction.

Consider a particle of mass $m$ and charge $q$ moving in an electromagnetic
field in a Minkowski's space $(x_1,x_2,x_3,x_4=ict)$, the 4-vector velocity
of the particle is denoted by $u_\mu $ , the 4-vector potential of the
electromagnetic field is denoted by $A_\mu $ , where and below we use Greek
letters for subscripts that range from 1 to 4. We write a theorem to
specify our argument.

\textbf{Theorem}\textit{: No mater how to move or when to move in the
Minkowski's space, the motion of the particle is governed by a potential
function }$\Phi $\textit{\ as}

\begin{equation}
mu_\mu +qA_\mu =\partial _\mu \Phi  \label{p1}
\end{equation}
\textit{For applying Eq.(\ref{p1}) to specific applications, we set }
$\Phi =-i\kappa \psi $\textit{, then Eq.(\ref{p1}) is
rewritten as}

\begin{equation}
(mu_\mu +qA_\mu )\psi =-i\kappa \partial _\mu \psi  \label{p2}
\end{equation}
\textit{the coefficient }$\kappa $\textit{\ is subject to the interpretation of }$%
\psi $.

Eq.(\ref{p1}) was obtained in the author's previous paper\cite{Cui114}, here
we shall not discuss its deduction, conversely, shall discuss how to use
it and reveal its relation with the Planck's constant.

There are three mathematical properties of $\psi $ worth recording here.
First, if there is a path $l_i$ joining initial point $x_0$ to final point $%
x $, then

\begin{equation}
\psi _i=e^{\frac i\kappa \int\nolimits_{x_0(l_i)}^x(mu_\mu +qA_\mu )dx_\mu }
\label{p3}
\end{equation}
Second, the integral of Eq.(\ref{p3}) is independent of the choice of
path. Third, the superposition principle is valid for $\psi _i$, i.e., if
there are $N$ paths from $x_0$ to $x$, then

\begin{equation}
\psi =\sum\limits_i^N\psi _i  \label{p4}
\end{equation}

\begin{equation}
m\overline{u_\mu }=\sum\limits_i^Nmu_\mu \psi _i/\sum\limits_i^N\psi _i\quad
\label{p5}
\end{equation}

\begin{equation}
(m\overline{u_\mu }+qA_\mu )\psi =-i\kappa \partial _\mu \psi  \label{p6}
\end{equation}
where $m\overline{u_\mu }$ is average momentum.

To gain further insight into physical meanings of this theorem, we shall
discuss four applications.

\section{Two slit experiment}

As shown in Fig.1, suppose that the electron gun emits a burst of electrons
at $x_0$ at time $t=0$, the electrons arrive at the point $x$ on the screen at 
time $t$. There are two paths for the electron to go to the destination, according 
to our above theorem, $\psi $ is given by

\begin{figure}[ht]
\includegraphics[bb=110 540 310 720,clip]{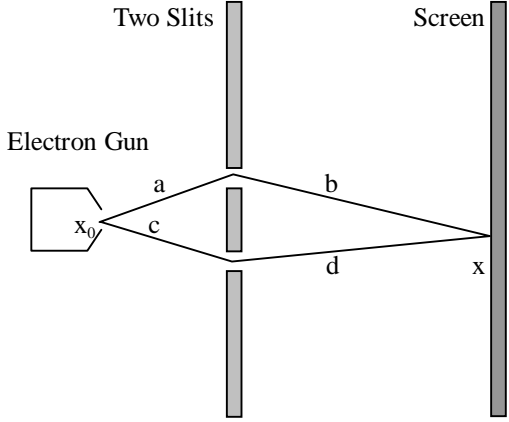}
\caption{A diffraction experiment in which electron beam from the gun
through the two slits to form a diffractin pattern at the screen.}
\end{figure}

\begin{equation}
\psi =e^{\frac i\kappa \int\nolimits_{x_0(l_1)}^x(mu_\mu )dx_\mu }+e^{\frac
i\kappa \int\nolimits_{x_0(l_2)}^x(mu_\mu )dx_\mu }  \label{p7}
\end{equation}
where we use $l_1$ and $l_2$ to denote the paths $a+b$ and $c+d$ respectively.
Multiplying Eq.(\ref{p7}) by its complex conjugate gives

\begin{eqnarray}
W&=&\psi (x)\psi ^{*}(x) \nonumber \\
&=&2+e^{\frac i\kappa \int\nolimits_{x_0(l_1)}^x(mu_\mu
)dx_\mu -\frac i\kappa \int\nolimits_{x_0(l_2)}^x(mu_\mu )dx_\mu }  \nonumber
\\
&&+e^{\frac i\kappa \int\nolimits_{x_0(l_2)}^x(mu_\mu )dx_\mu -\frac i\kappa
\int\nolimits_{x_0(l_1)}^x(mu_\mu )dx_\mu }  \nonumber \\
&=&2+2\cos [\frac 1\kappa \int\nolimits_{x_0(l_1)}^x(mu_\mu )dx_\mu 
\nonumber \\
&&-\frac 1\kappa \int\nolimits_{x_0(l_2)}^x(mu_\mu )d_\mu ]  \nonumber \\
&=&2+2\cos [\frac p\kappa (l_1-l_2)]  \label{p8}
\end{eqnarray}
where $p$ is the momentum of the electron. We find a typical interference pattern with constructive interference when $l_1-l_2$ is an integral multiple of $\kappa /p$, and destructive interference when it is a half integral multiple. This kind of experiment has been
done a long age, no mater what kind of particle, the comparision of the experiment to Eq.(\ref{p8}) leads to two consequences: (1) the complex function $\psi $ is found to be 
probability amplitude, i.e., $\psi (x)\psi ^{*}(x)$ expresses the probability of finding a particle at location $x$ in the Minkowski's space. (2) $\kappa $ is the Planck's constant.

\section{The Aharonov-Bohm effect}

Let us consider the modification of the two slit experiment, as shown in
Fig.2. Between the two slits there is located a tiny solenoid S, designed so
that a magnetic field perpendicular to the plane of the figure can be
produced in its interior. No magnetic field is allowed outside the solenoid,
and the walls of the solenoid are such that no electron can penetrate to the
interior. Like Eq.(\ref{p7}), the amplitude $\psi $ is given by

\begin{equation}
\psi =e^{\frac i\kappa \int\nolimits_{x_0(l_1)}^x(mu_\mu +qA_\mu )dx_\mu
}+e^{\frac i\kappa \int\nolimits_{x_0(l_2)}^x(mu_\mu +qA_\mu )dx_\mu }
\label{p9}
\end{equation}
and the probability is given by

\begin{eqnarray}
W&=&\psi (x)\psi ^{*}(x) \nonumber \\
&=&2+e^{\frac i\kappa \int\nolimits_{x_0(l_1)}^x(mu_\mu
+qA_\mu )dx_\mu -\frac i\kappa \int\nolimits_{x_0(l_2)}^x(mu_\mu +qA_\mu
)dx_\mu }  \nonumber \\
&&+e^{\frac i\kappa \int\nolimits_{x_0(l_2)}^x(mu_\mu +qA_\mu )dx_\mu -\frac
i\kappa \int\nolimits_{x_0(l_1)}^x(mu_\mu +qA_\mu )dx_\mu }  \nonumber \\
&=&2+2\cos [\frac p\kappa (l_1-l_2)+\frac 1\kappa
\int\nolimits_{x_0(l_1)}^xqA_\mu dx_\mu  \nonumber \\
&&-\frac 1\kappa \int\nolimits_{x_0(l_2)}^xqA_\mu dx_\mu ]  \nonumber \\
&=&2+2\cos [\frac p\kappa (l_1-l_2)+\frac 1\kappa \oint_{(l_1+\overline{l_2}%
)}qA_\mu dx_\mu ]  \nonumber \\
&=&2+2\cos [\frac p\kappa (l_1-l_2)+\frac{q\phi }\kappa ] \label{p10}
\end{eqnarray}
where $\overline{l_2}$ denotes the inverse path to the path $l_2$, $\phi $
is the magnetic flux that passes through the surface between the paths $l_1$
and $\overline{l_2}$, and it is just the flux inside the solenoid. 

\begin{figure}[ht]
\includegraphics[bb=110 540 310 720,clip]{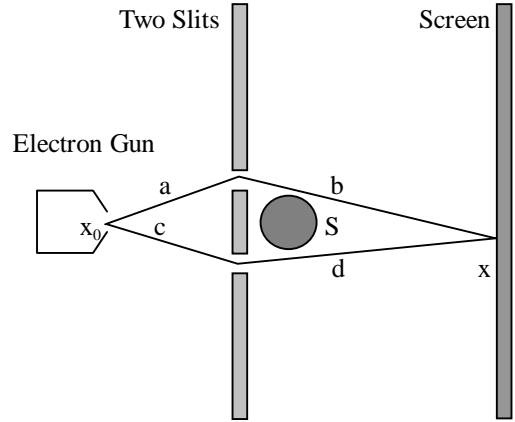}
\caption{A diffraction experiment with adding a solenoid.}
\end{figure}

Now, constructive (or destructive) interference occurs when

\begin{equation}
\frac p\kappa (l_1-l_2)+\frac{q\phi }\kappa =2\pi n\quad (or\quad n+\frac 12)
\label{p11}
\end{equation}
where $n$ is an integer. When $\kappa $ takes the value of the Planck's
constant, we know that this effect is just the Aharonov-Bohm effect which
was shown experimentally in 1960.

\section{The hydrogen atom}

The hydrogen atom is one of the few physically significant
quantum-mechanical systems for which an exact solution can be found and the
theoretical predictions compared with experiment.

Rutherford's model of a hydrogen atom consists of a nucleus made up of a
single proton and of a single electron outside the nucleus, the electron
moves in an orbit about the nucleus. Here we consider two points denoted by $%
x_0$ and $x$ in the orbit, and two paths $l_1$ and $l_2$ from $x_0$ to $x$
along different directions, as shown in Fig.3. Then, according to our above theorem, the probability amplitude $\psi $ is given by

\begin{figure}[ht]
\includegraphics[bb=110 595 310 705,clip]{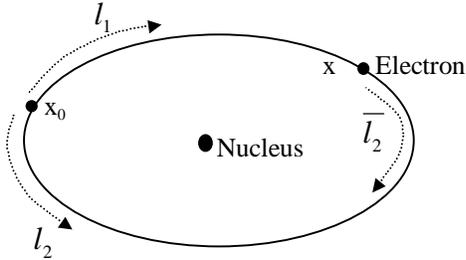}
\caption{The electron moves in an orbit about the necleus.}
\end{figure}

\begin{equation}
\psi =e^{\frac i\kappa \int\nolimits_{x_0(l_1)}^x(mu_\mu +qA_\mu )dx_\mu
}+e^{\frac i\kappa \int\nolimits_{x_0(l_2)}^x(mu_\mu +qA_\mu )dx_\mu }
\label{p12}
\end{equation}
and the probability is given by 
\begin{eqnarray}
W&=&\psi (x)\psi ^{*}(x) \nonumber \\
&=&2+2\cos [\frac 1\kappa \oint_{(l_1+\overline{l_2}%
)}(mu_\mu +qA_\mu )dx_\mu ]  \nonumber \\
&=&2+2\cos [\frac 1\kappa \oint_{(l_1+\overline{l_2})}(mu_k)dx_k]  \nonumber
\\
&=&2+2\cos [\frac 1\kappa \oint_{(l_1+\overline{l_2})}p_kdx_k]  \label{p13}
\end{eqnarray}
where $k=1,2,3$, $p_k=mu_k$. For the stationary states, the integral about
time will be automatically eliminated because the probability should be stable.
The probability of the electron at every point in the orbit should be the
same because these points in the orbit  are equivalent, this leads to

\begin{equation}
\oint_{(orbit)}p_kdx_k=2\pi \kappa n  \label{p14}
\end{equation}

When $\kappa =\hbar $, Eq.(\ref{p14}) is just the Bohr-Somerfeld
quantization rule for the hydrogen atom.

The probability of the electron outside the orbit should vanish, in where
the momentum of the electron should become imaginary.

\section{The motion of particle in a potential well}

Let us now restrict ourselves to one dimensional well. We choose point $x_0$
to locate at the left turning point and $x$ at arbitrary point in the well,
as shown in Fig.4, likewise, there are two paths $l_1$ and $l_2$ from $x_0$
to $x$ to correspond to ''coming'' ($l_1$) and ''back'' ($\overline{l_2}$)
for the particle motion, like Eq.(\ref{p12}) and (\ref{p13}), we obtain the
probability as

\begin{figure}[ht]
\includegraphics[bb=110 550 310 720,clip]{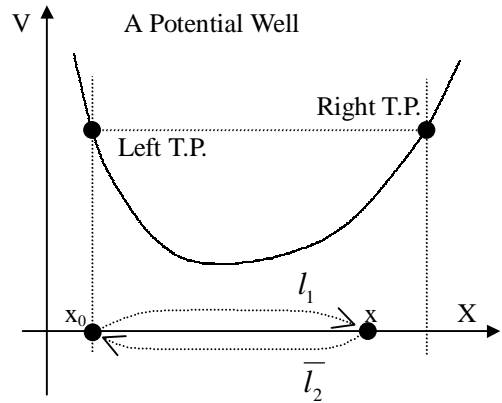}
\caption{The motion of particle in a potential well. The left and right
turning points are indicated on the potential well. }
\end{figure}

\begin{eqnarray}
W&=&\psi (x)\psi ^{*}(x) \nonumber \\
&=&2+2\cos [\frac 1\kappa \oint_{(l_1+\overline{l_2}%
)}(mu_\mu +qA_\mu )dx_\mu ]  \nonumber \\
&=&2+2\cos [\frac 1\kappa \oint_{(l_1+\overline{l_2})}pdx]  \label{p15}
\end{eqnarray}
The integral about time vanishes for the stationary state. The probability
has a distribution in the well, but it will vanish at the right turning
point for satisfying boundary condition, this leads to

\begin{equation}
\oint_{}pdx=2\pi \kappa (n+\frac 12)  \label{p16}
\end{equation}
where the integral is evaluated over one whole period of classical motion,
from the left turning point to the right and back. We again meet the
Bohr-Sommerfeld quantization rule for the old quantum theory when we take $%
\kappa =\hbar $, although it was originally written in the form of Eq.(\ref{p14}) in 
1915 due to A. Sommerfeld and W. Wilson.

\section{Discussion}

The above formulation based on the theorem of Eq.(\ref{p1}) is successful to the quantum
mechanics, but we emphasize that Eq.(\ref{p1}) is essentially different from
the Schrodinger's equation. In the author's previous paper we have proved
that we can derive the Schrodinger's equation from our Eq.(\ref{p1}),
inversely we can not obtain Eq.(\ref{p1}) from the Schrodinger's equation.

We always assume that the path integral about time vanishes for stationary
state, because we always investigate stable experimental phenomena. If we
can be equipped to investigate dynamic processes, the path integral about time
will display its effects.

\section{Conclusion}

The Planck's constant is an fundamental constant which can be well defined
in the theorem of Eq.(\ref{p1}).


\begin{thebibliography}{9}
\bibitem{Cui114}  H. Y. Cui, eprint, quant-ph/0102114,(2001).

\bibitem{Harris}  E. G. Harris, Introduction to Modern Theoretical Physics,
Vol.1\&2, (John Wiley \& Sons, USA, 1975).

\bibitem{Schiff}  L. I. Schiff, Quantum Mechanics, third edition, (McGraw-Hill, USA, 1968).

\bibitem{Sakurai}  J. J. Sakurai, Modern Quantum Mechanics,
(Benjamin/Cummings, USA, 1985).

\bibitem{Cui89}  H. Y. Cui, College Physics, \textbf{4}, 13(1989).

\bibitem{Cui073}  H. Y. Cui, eprint, physcis/0102073, (2001).
\end{thebibliography}
\end{document}